\documentstyle[aps,prl,preprint]{revtex}
\def\la{\mathrel{\mathpalette\fun <}}

\def\fun#1#2{\lower3.6pt\vbox{\baselineskip0pt\lineskip.9pt
        \ialign{$\mathsurround=0pt#1\hfill##\hfil$\crcr#2\crcr\sim\crcr}}}
\def\l{\lambda}
\begin{document}
\begin{titlepage}
\vspace*{-62pt}
\begin{flushright}
DART-HEP-94/01\\
March 1994
\end{flushright}

\vspace{0.75in}
\centerline{\bf DYNAMICS OF WEAK FIRST ORDER PHASE TRANSITIONS$^{\ast}$}
\vskip 1.0cm
\centerline{Marcelo Gleiser}

\vskip 1.5 cm
\centerline{\it Department of Physics and Astronomy, Dartmouth College,
Hanover, NH 03755}

\vskip 1.5 cm
\centerline{\bf Abstract}
\begin{quote}
{The dynamics of weak vs. strong first order phase transitions
is investigated numerically for 2+1 dimensional scalar field models.
It is argued that the change from
a weak to a strong transition is itself a (second order) phase transition,
with the order
parameter being the equilibrium fractional population
difference between the two phases
at the critical temperature,
and the control parameter being the coefficient of the cubic coupling
in the free-energy density.
The critical point is identified, and
a power law controlling the relaxation dynamics at this point is obtained.
Possible applications are briefly discussed.}

\vspace{48pt}
$\ast$ electronic mail: gleiser@peterpan.dartmouth.edu

\end{quote}
\end{titlepage}
\def\beq{\begin{equation}}
\def\eeq{\end{equation}}
\def\ba{\begin{eqnarray}}
\def\ea{\end{eqnarray}}
\def\re#1{[{\ref{#1}]}}

\def\mpl{{m_{Pl}}}
\def\x{{\bf x}}
\def\p{\phi}
\def\F{\Phi}
\def\s{\sigma}
\def\a{\alpha}
\def\d{\delta}
\def\t{\tau}
\def\r{\rho}

\vskip 1cm
\vspace{16pt}

Of the many interesting possibilities raised by primordial phase
transitions \re{CPT},
the generation of the baryon number of the Universe during the electroweak
transition has been extensively investigated following the seminal work
of Kuzmin, Rubakov, and Shaposhnikov \re{KRS}. For the purpose of this paper,
the important aspect of the electroweak phase transition is that it is, in
most scenarios proposed so far,
a first order
phase transition. And, at least within the context of the standard model of
particle physics, the transition is very possibly a weak one; the standard
computation for nucleation of critical bubbles (see Refs.
[\ref{CH}~-~\ref{FINITETDECAY}])
shows that the thin-wall
approximation fails and that the bubbles are rather thick \re{THICK}.

The possibility that the electroweak transition could be weakly first order
has led Gleiser and Kolb (GK) to propose a novel mechanism by
which such transitions
evolve \re{GK}.
The transition would be characterized by
a substantial phase mixing as the critical temperature $T_c$
({\it i.e.} when the
two phases are degenerate) is approached from
above, followed by domain coarsening below $T_c$. (The slow cooling is
provided by the expansion of the Universe.)
GK modelled the dynamics of this phase mixing by estimating
the fraction of the volume occupied by sub-critical (correlation volume)
thermal fluctuations of
each phase as a function of the temperature.
They neglected the fact that these sub-critical fluctuations were
unstable and thus found that the system would be equally populated by the
two phases as it reached $T_c$.
The results of Ref. \re{GG} indicate that
GK are at least qualitatively correct; there will be a regime in which the
transition is
weak enough that considerable phase mixing occurs even above $T_c$. (It is of
course possible, although not directly relevant here,
that this interesting regime lies beyond the validity of the
perturbative evaluation of the electroweak effective potential.
Presently this question does not appear to be resolved \re{EFFECTIVEPOT}.)

Due to the complex nonequilibrium nature of the system, any analytical approach
is bound to be severely limited.
The need for a numerical investigation of this question
is clear. This need is even more justified by noting that several of the
{\it gross} features
of the electroweak transition may appear in other unrelated physical systems
such as nematic liquid crystals and certain magnetic materials.
Moreover, numerical
simulations of first-order transitions in the context of field theories
(as opposed to discrete Ising models \re{ISINGNUM}) are scarse.
Recent work has shown that
the effective nucleation barrier is accurately predicted by homogeneous
nucleation
theory in the context of 2+1-dimensional classical field theory \re{AG}. These
results were obtained for strong transitions, in which the nucleation barrier
$B$ was large. Nucleation was made possible due to the fairly high
temperatures
used in the simulations. (Recall that the decay time is proportional to
${\rm exp}(B/T)$.)

In order to study how the weakness of the transition will affect its dynamics,
the homogeneous part of the (coarse-grained Helmholtz) free-energy density
is written as
\beq
U(\p ,T)={a\over 2}\left (T^2-T_2^2\right )\p^2-
{{\alpha}\over 3}T\p^3 +{{\l}\over 4}
\p^4 \:.
\label{e:freen}
\eeq
This free-energy density resembles the finite-temperature
effective potential
used in the description of the electroweak transition, where $\alpha$ is
determined by the masses of the gauge bosons and $T_2$ is the spinodal
instability temperature
\re{KRS}. In the electroweak case the order parameter
is the magnitude of the Higgs field and the effective potential is
obtained after integrating out the gauge and fermionic degrees of freedom.
Here, we will not be concerned with the
limits of validity of the perturbative effective potential.
The goal is to explore the possible dynamics of a
transition with free-energy density
given by Eq.~\ref{e:freen}, and use the results as
suggestive of the behavior in the electroweak case.
This free-energy density
is also similar
to the de Gennes-Ginsburg-Landau free energy
(with the elastic constants set to zero) used in
the study of the isotropic-nematic transition in liquid crystals \re{LIQCRYS}.
This transition is known to be weakly first-order;
departures from the mean field prediction for the behavior of the
correlation length were detected as the degeneracy temperature is approached
from above, signalling the presence of ``pre-transitional phenomena'',
due to long-wavelength fluctuations observed by
light-scattering experiments \re{STINSON}.

It proves convenient to introduce dimensionless variables,
${\tilde x}=x\sqrt{a}T_2,~{\tilde t}=t\sqrt{a}T_2,~X=\p/\sqrt{T_2},
{\rm and}~\theta=
T/T_2$, so that we can write the Hamiltonian as
\beq
{{H[X]}\over {\theta}}={1\over {\theta}}\int d^2{\tilde x}\left [
{1\over 2}\mid {\tilde \bigtriangledown} X\mid^2 +
{1\over 2}\left (\theta^2 -1
\right )X^2 -{{
{\tilde \alpha}}\over 3}X^3+{{{\tilde \l}}\over 4}X^4\right ] \:,
\label{e:hamilton}
\eeq
where ${\tilde \a}=\a/(a\sqrt{T_2})$, and ${\tilde \l}=\l/(aT_2)$. (From
now on the tildes will be dropped unless a new quantity is introduced.)
For temperatures above $\theta_1=(1-\a^2/4\l)^{-1/2}$ there is only one
minimum at $X=0$. At $\theta=\theta_1$ an inflection point appears at
$X_{\rm inf}=\a\theta_1/2\l$. Below $\theta_1$ the inflection point separates
into a maximum and a minimum given by $X_{\pm}={{\a\theta}\over {2\l}}\left [
1\pm \sqrt{1-4\l\left (1-1/\theta^2\right )/\a^2}\right ]$. At the critical
temperature $\theta_c=(1-2\a^2/9\l)^{-1/2}$ the two minima, at $X_0=0$ and
$X_+$ are degenerate. Below $\theta_c$ the minimum at $X_+$ becomes the global
minimum and the $X_0$-phase becomes metastable. Finally, at $\theta=1$ the
barrier between the two phases disappears.

In order to study numerically
the approach to equilibrium at a given temperature
$\theta$, the coupling of the order
parameter $X$ with the thermal bath will be modelled by a Markovian Langevin
equation,
\beq\label{e:langevin}
{\partial^2X\over\partial t^2} = \bigtriangledown^2X -
{\tilde \eta} {\partial X\over
\partial t}
- {\partial U(X,\theta) \over \partial X} + {\tilde \xi}(x,t)~~,
\eeq
where ${\tilde \eta}=\eta/\sqrt{a}T_2$ is the dimensionless
viscosity coefficient,
and ${\tilde \xi}=\xi/aT_2^{5/2}$ is the dimensionless
stochastic noise with vanishing mean, related to
$\eta$ by the fluctuation-dissipation theorem,
$\langle \xi(\vec x,t)\xi( \vec x',t')\rangle =
  2\eta T\delta(t-t')\delta^2(\vec x - \vec x')$.
The viscosity coefficient was set to unity in all simulations.
The lattice spacing was also set to unity in all simulations.
It turns out that in all cases of interest here
the mean-field correlation length $\xi^{-2}_{{\rm cor}} = U^{\prime \prime}
(X_0,\theta)$
 will be sufficiently larger than unity to justify this choice.
In future work it would
be interesting to see how to generalize the lattice renormalization conditions
obtained in Ref. \ref{AG} for temperature independent potentials to the
situation studied here.
The Langevin equation was integrated using the fifth-degree Nordsiek-Geer
algorithm, which allows for fast integration with high numerical accuracy
\re{NORD}. The time step used was $\delta t=0.2$, and results
were obtained with a square lattice with $L=64$. (Comparison with $L=40$ and
$L=128$ produced negligible differences for our present purposes.)
No dependence of the
results was found on the time-step,
random noise generator, and random noise seed.

The strategy adopted was to study the behavior of the system given by Eq.
\re{e:hamilton} at the critical temperature when the two minima are
degenerate. The reason for this choice of temperature is simple. If at
$\theta_c$ most of the system is found in the $X=0$ phase then as the
temperature drops below $\theta_c$, one expects homogeneous nucleation to
work; the system is well-localized in its metastable phase. This is what
happens when a system is rapidly cooled below its critical temperature
(rapid quench), so that it finds itself trapped in the metastable state. The
large amplitude fluctuations which will eventually appear and grow
are the nucleating
bubbles.
If, at $\theta_c$ one finds a large probability for the system to be in the
$X_+$-phase, then considerable phase-mixing is occuring and homogeneous
nucleation should not be accurate in describing the transition. Large amplitude
fluctuations are present in the system before it is quenched to
temperatures below $\theta_c$.
For definiteness call the two phases the 0-phase and the +-phase.
The phase distribution of the system can be measured if the idea of fractional
area (volume in three dimensions) is introduced. As the field evolves according
to Eq. \ref{e:langevin},
one counts how much of the total area of the lattice belongs to the 0-phase
with $X\leq X_-$ ({\it i.e.} to the left of the maximum),
and how much belongs to the +-phase with $X>X_-$ ({\it i.e.} to the right of
the maximum). Dividing by the total area one obtains the fractional area in
each phase, so that $f_0(t) + f_+(t)=1$, indepedently of $L^2$.

The system is prepared initially in the 0-phase, $f_0(0)=1$ and $f_+(0)=0$.
Thus, the area-averaged value of the order parameter, $\langle X\rangle (t)=
A^{-1}\int X dA$ is initially zero. The coupling with the thermal bath will
induce fluctuations about $X=0$. By keeping $\l=0.1$ fixed, the dependence of
$f_0(t),~f_+(t)$, and  $\langle X\rangle (t)$ on the value of
$\a$ can be measured. Larger values of $\a$ imply
stronger transitions. This is clear from the expression for $\theta_c$ which
approaches unity as $\a\rightarrow 0$.
(In the electroweak case, the
same argument applies, as what is relevant is the ratio $\a^2/\l$; $\a$ is
fixed but $\l$ increases as the Higgs mass increases.) The results are
shown in Figure 1 for several values of $\a$ between $\a=0.3$ and $\a=0.4$.
Each one of these curves is the result of
averaging over 200 runs. The two important features here are the final
equilibrium fraction in each phase and the equilibration time-scale.
The approach to equilibrium can be fitted {\it at all times} to an exponential,

\beq
f_0(t) = \left(1 - f_0^{{\rm EQ}}\right ){\rm exp}\left [
-\left (t/\t_{{\rm EQ}}
\right )^{\sigma}\right ] + f_0^{{\rm EQ}}\:,
\label{e:equil}
\eeq
where $f_0^{{\rm EQ}}$ is the final equilibrium fraction and $\t_{{\rm EQ}}$
is the equilibration time-scale. In Table 1 the
values of $\t_{{\rm EQ}}$ and $\sigma$ are listed for different values of
$\a$. Note that the slot for $\a=0.36$ is empty. For this value of $\a$ the
approach to equilibrium cannot be fitted at all times to an exponential;
however, at large times it can be fitted
to a power law,
\beq
f_0(t)\mid_{\a=0.36}~ \propto~  t
^{-k} \:,
\label{e: powerlaw}
\eeq
where $k$ is the critical exponent controlling the approach to equilibrium.
A good fit is obtained for $k = 0.25 ~(\pm 0.01) $
as shown in Figure 2. Note that this is not the same as the dynamical
critical exponent $z\simeq 2$, defined as $\t_{{\rm EQ}} = \xi_{{\rm cor}}^z$.

The fact that there is a critical slowing down of the dynamics for $\a
\simeq 0.36$
is indicative of the presence of a second order
phase transition near $\a\simeq 0.36$. Similar behavior has been found
in liquid crystals in the neighborhood of the isotropic-nematic transition
\re{LIQCRYSRELAX}, and is typical of ferromagnetic materials near the Curie
temperature.
This transition reveals itself in a striking way if we define as an order
parameter the equilibrium fractional difference $\Delta F_{{\rm EQ}}$,
\beq
\Delta F_{{\rm EQ}} = f_0^{{\rm EQ}} - f_+^{{\rm EQ}} \:.
\label{e:fbar}
\eeq
In Figure 3  $\Delta F_{{\rm EQ}}$
is plotted as a function of $\a$. Clearly, there is a marked change in the
behavior of the system around $\a=\a_c\simeq 0.36$. This curve is essentially
identical
to numerical results for the magnetization as a
function of temperature in
Ising models; the rounding is due to finite size effects. (See Fig. 2a in
\re{ISINGNUM}.)
For $\a < \a_c$ the
fractional area occupied by both phases in equilibrium
is practically the same at 0.5. There
is considerable mixing of the two phases, with the system unable to
distinguish between them.
One may call this phase the symmetric phase with
respect to the order parameter  $\Delta F_{{\rm EQ}}$. For $\a > \a_c$ there
is a clear distinction between the two phases, with the +-phase being sharply
suppressed. This may be called the broken-symmetric phase.
As a consequence of this behavior
a very clear distinction between a strong and weak transition is possible.
A strong transition has $\a > \a_c$ so that the system is dominated by the
0-phase at $\theta_c$. For a weak transition neither phase clearly dominates
and, as argued above, the dynamics should be quite different from
the usual nucleation mechanism.

In order to understand the reason for the sharp change of behavior
of the system near $\a_c$, in Figure 4 the equilibrium
area-averaged order paramemeter $\langle X\rangle_{{\rm EQ}}$ and the
inflection point $X_{{\rm inf}} = {{\a\theta}\over {3\l}}\left [1-\sqrt{1-
3\l(1-1/\theta^2)/\a^2}\right ]$, are shown as a function of $\a$. Also shown
is the rms amplitude of correlation-size fluctuations $X_{{\rm rms}}^2 =
\langle X^2\rangle_T - \langle X\rangle^2_T $, where
$\langle \cdots \rangle_T$
is the normalized thermal average with probability distribution $P[X_{sc}
] ={\rm
exp}\left [-F[X_{sc}]/\theta\right ]$. $F[X_{sc}]$ is the free energy of a
gaussian-shape sub-critical fluctuation.
For details see Ref. \re{GR}. It is clear
from this figure that the transition from weak to strong occurs as
$\langle X\rangle_{{\rm EQ}}$ drops below  $X_{{\rm inf}}$.
This result can be interpreted as an effective Ginzburg criterion
for the weak-to-strong transition. It matches quite
well the fact that the critical slowing down occurs for $\a\simeq 0.36$.
This result is in qualitative agreement with the study of Langer {\it et al.}
contrasting the onset of nucleation vs. spinodal decomposition
for binary fluid and solid solutions
\re{LANGER II}, where it was found that the transition between the two
regimes occurs roughly at the spinodal ({\it i.e.} at the inflection point).
Even though $X_{{\rm rms}}$ drops below  $X_{{\rm inf}}$ for a smaller
value of $\a$, being a much less computer intensive quantitity to obtain,
it should serve as a rough indicator of the weak-to-strong transition.

The present work raises many questions for future investigation. Apart
from investigating the 3+1-dimensional case,
and obtaining the critical exponent for the order parameter (as well
as more accurate values for $\a_c$, $\t_{{\rm EQ}}$ and $k$) using
finite-size scaling techniques \re{ISINGNUM},
$\Delta F_{{\rm EQ}} \propto (\a - \a_c)^{\beta}$ ($\beta=1/8$ for the $d=2$
Ising model, and $\beta=1/2$ for mean field),
it should
be interesting to test if this behavior could be observed in the
laboratory. A possible system would be an Ising
magnetic film in the absence of an external field, heated to just above its
{\it measured} Curie point. For the model studied here, this would correspond
to $\a \la \a_c$.

\acknowledgements
I am grateful to W. E. Lawrence for many enlightening comments and to
M. Alford, H. M\"uller, and Rudnei Ramos
for useful discussions. I also thank F. F. Abraham for his guidance with
the numerical work
during the early stages of this investigation.
This work is partially supported by a National Science Foundation grant
No. PHYS-9204726.

\references
\begin{enumerate}

\item\label{CPT} E. W. Kolb and M. S. Turner, {\it The Early Universe},
Addison-Wesley (1990).

\item\label{KRS} V. A. Kuzmin, V. A. Rubakov,
and M. E. Shaposhnikov, {\em Phys. Lett.} {\bf 155B}, 36 (1985).
For a recent review see, A. Cohen, D. Kaplan, and A. Nelson, Ann. Rev.
Nucl. Part. Sci. {\bf 43}, 27 (1993).

\item\label{CH}
For a review of homogeneous nucleation see,
J. D. Gunton, M. San Miguel,
and P. S. Sahni, in
{\it Phase Transitions and Critical Phenomena}, edited by C. Domb and J. L.
Lebowitz (Academic, London, 1983), Vol. 8.

\item\label{COLEMAN} S. Coleman, Phys. Rev. {\bf D15}, 2929 (1977); C. Callan
and S. Coleman, Phys. Rev. {\bf D16}, 1762 (1977).

\item\label{FINITETDECAY}  A. D. Linde, Nucl. Phys. {\bf B216}, 421 (1983);
[Erratum:
{\bf B223}, 544 (1983)]; for recent work, see e.g.,
L. P. Csernai and J. I. Kapusta, Phys. Rev. {\bf D46},
1379 (1992); M. Gleiser, G. Marques, and R. Ramos, Phys. Rev. {\bf D48},
1571 (1993); D. Brahm and C. Lee, Caltech preprint No. CALT-68-1881 (1993).

\item\label{THICK} See, e.g., B. Liu, L. McLerran, and N. Turok, Phys. Rev.
{\bf D46}, 2668 (1992); M. Dine and S. Thomas, Santa Cruz preprint No.
SCIPP 94/01.

\item\label{GK} M. Gleiser and E. W. Kolb, Phys. Rev. Lett. {\bf 69},
1304 (1992); Phys. Rev. {\bf D48}, 1560 (1993).

\item\label{GG} M. Gleiser and G. Gelmini, in press, Nucl. Phys. B. Criticism
to the GK approach can be found in this reference.

\item\label{EFFECTIVEPOT} P. Arnold and L. Yaffe , Univ. of Washington
preprint No. UW/PT-93-24, and refs. therein.

\item\label{ISINGNUM} D. P. Landau in {\it Finite Size Scaling and
Numerical Simulations of Statistical Systems}, edited by V. Privman
(World Scientific, Singapore, 1990).

\item\label{AG} M. Alford and M. Gleiser, Phys. Rev. {\bf D48}, 2838 (1993);
O. T. Valls and G. F. Mazenko, Phys. Rev. {\bf B42},
6614 (1990).

\item\label{LIQCRYS}
S. Chandrasekhar,
{\it Liquid Crystals}, (Cambridge University Press, Cambridge
[Second edition, 1992]).

\item\label{STINSON} T. W. Stinson and
J. D. Litster, {\it Phys. Rev. Lett.} {\bf 25}, 503 (1970); H. Zink and
W. H. de Jeu, {\it Mol. Cryst. Liq. Cryst.} {\bf 124}, 287 (1985).

\item\label{NORD} A. Nordsiek, Math. Comp. {\bf 16}, 22 (1962).

\item\label{LIQCRYSRELAX} F. W. Deeg and M. D. Fayer, Chem. Phys. Lett.
{\bf 167}, 527 (1990); J. D. Litster and T. W. Stinson, J. App. Phys.,
{\bf 41}, 996 (1970).

\item\label{GR} M. Gleiser and R. Ramos, Phys. Lett. {\bf B300}, 271 (1993).

\item\label{LANGER II} J. Langer, Physica {\bf 73}, 61 (1974);
J. Langer, M. Bar-on, and H. Miller, Phys. Rev. {\bf A11}, 1417 (1975). Note
that these authors studied transitions with a conserved order parameter, and
thus with slower dynamics.

\end{enumerate}

\listoffigures

Figure 1. The approach to equilibrium for several values of $\a$.
         \\

Figure 2. Fitting $f_0(\theta_c)$ to a power law at large times for
$\a=0.36$.\\

Figure 3. The fractional equilibrium population difference $\Delta
F_{{\rm EQ}}$ as a function of $\a$.\\

Figure 4. Comparison between area-averaged field and location of the
inflection point as a function of $\a$. Also shown are the location of the
barrier, $X_{{\rm MAX}}$ and the rms fluctuation $X_{{\rm rms}}$.

\listoftables

Table 1. The values of the equilibration time-scales and the exponents for
the exponential fit of Eq. \ref{e:equil}
for several values of $\a $. Also shown are the
equilibrium fractions $f_0(\theta_c)$ and $f_+(\theta_c)$. Uncertainties are
in the last digit.

\vfill\eject

\centerline {\bf TABLE 1}
\vspace{2.in}

\begin{center}
\begin{tabular}{|c|c|c|c|c|}\hline\hline
$\a $ & $\t_{{\rm EQ}}$ & $\sigma $ & $f_0(\theta_c)$ & $f_+(\theta_c)$ \\
\hline\hline
 0.30 & 21.0 & 0.80 & 0.505 & 0.495\\
 0.33 & 40.0 & 0.80 & 0.514 & 0.486\\
 0.35 & 75.0 & 0.60 & 0.525 & 0.475\\
 0.36 & --   & --   & 0.580 & 0.420\\
 0.37 & 25.0 & 0.65 & 0.800 & 0.200 \\
 0.38 & 15.0 & 0.80 & 0.870 & 0.130\\
 0.40 & 5.0  & 1.0  & 0.937 & 0.063\\
\hline
\end{tabular}
\end{center}

\end{document}